\documentclass[11pt,onecolumn]{IEEEtran}

\usepackage{amsmath}   
\RequirePackage{amsthm,amssymb,amsfonts}
\usepackage{hyperref}
\hypersetup{colorlinks=true,pdftitle="Ruzsa",pdftex}

\newtheorem{thm}{Theorem}
\newtheorem{conj}{Conjecture}
\newtheorem{cor}{Corollary}
\newtheorem{lem}{Lemma}
\newtheorem{defn}{Definition}
\newtheorem{rmk}{Remark}
\newtheorem{ex}{Example}

\usepackage{mathtools}

\DeclarePairedDelimiter{\floor}{\lfloor}{\rfloor}

\newcommand{\bc}{\begin{center}}
\newcommand{\ec}{\end{center}}
\newcommand{\bt}{\begin{tabular}}
\newcommand{\et}{\end{tabular}} 
\newcommand{\bea}{\begin{eqnarray}}
\newcommand{\eea}{\end{eqnarray}}

\newcommand{\ba}{\begin{array}}
\newcommand{\ea}{\end{array}}

\def\be{\begin{eqnarray}}
\def\ee{\end{eqnarray}}
\def\ben{\begin{eqnarray*}}
\def\een{\end{eqnarray*}}

\newcommand{\ra} {\rightarrow}

\newcommand{\plusE}{\mbox{$ \;\stackrel{E}{+}\; $}}

\newcommand{\nth}{\frac{1}{n}}

\newcommand{\RL}{{\mathbb R}}
\newcommand{\ZZ}{{\mathbb Z}}

\newcommand{\calF}{\mbox{${\cal F}$}}

\newcommand{\Nat}{\mathbb{N}}

\def\sq{$\Box$}

\def\qed{\ifmmode\sq\else{\unskip\nobreak\hfil
\penalty50\hskip1em\null\nobreak\hfil\sq
\parfillskip=0pt\finalhyphendemerits=0\endgraf}\fi\par\medbreak}

\newsavebox{\junk}
\savebox{\junk}[1.6mm]{\hbox{$|\!|\!|$}}

\def\det{{\mathop{\rm det}}}

\def\half{{\mathchoice{\textstyle \frac{1}{2}}%
{\frac{1}{2}}%
{\hbox{\tiny $\frac{1}{2}$}}%
{\hbox{\tiny $\frac{1}{2}$}} }}

 \def\eq#1/{(\ref{#1})}

\def\eq#1/{(\ref{e:#1})}

\newcommand{\bbP}{\mathbb{P}}

\newcommand{\calN}{\mathcal{N}}
\newcommand{\calG}{\mathcal{G}}

\newcommand{\lam}{\lambda}

\newcommand{\vol}{\text{Vol}}

\def\E{{\bf E}}

\def\phi{\varphi}

\def\bee{\begin{eqnarray*}}
\def\ene{\end{eqnarray*}}

\begin{document}
%
\title{Entropy bounds on abelian groups\\
and the Ruzsa divergence}
\author{Mokshay~Madiman,~\IEEEmembership{Member,~IEEE} and~Ioannis~Kontoyiannis,~\IEEEmembership{Fellow,~IEEE}
\thanks{M. Madiman is with the Department of Mathematical Sciences,
University of Delaware, 501 Ewing Hall, Newark DE 19716, USA.
Email: {\tt madiman@udel.edu}}
\thanks{I.~Kontoyiannis is with the Department of Informatics, 
Athens University of Economics and Business, Patission 76, Athens 10434, Greece.
Email: \texttt{yiannis@aueb.gr}}
\thanks{Manuscript dates. 
M.~Madiman was supported by NSF CAREER grant DMS-1056996 and NSF grant CCF-1065494.
I.~Kontoyiannis was supported by the European Union (European Social Fund 
- ESF) and Greek national funds through the Operational Program 
``Education and Lifelong Learning'' of the National Strategic Reference 
Framework (NSRF) Research Funding Program 
``Thales - Investing in knowledge society through the European Social Fund.''
Parts of this paper were presented at the IEEE Information Theory Workshop 2013 in Seville, Spain,
and at the Eighth Workshop on Information Theoretic Methods in Science and Engineering (WITMSE 2015)
in Copenhagen, Denmark.
}
}
%
%
%
\markboth{Submitted to IEEE Transactions on Information Theory, 2015}{Madiman and Kontoyiannis}
%



\maketitle

\begin{abstract}
Over the past few years, a family of interesting
new inequalities for the entropies of sums and
differences of random variables has 
been developed by Ruzsa, Tao and others,
motivated by analogous results in additive combinatorics.
The present work 
extends these earlier results to the case of random 
variables taking values in $\RL^n$ or, more generally, in arbitrary
locally compact and Polish abelian groups. We isolate
and study a key quantity, the {\em Ruzsa divergence}
between two probability distributions,
and we show that its properties can be used 
to extend the earlier inequalities to the 
present general setting. The new 
results established include several variations
on the theme that the entropies of the sum 
and the difference of two independent random 
variables severely constrain each other.
Although the setting is quite general, the result 
are already of interest (and new) for random 
vectors in $\RL^n$. In that special case,
quantitative bounds are provided for the
stability of the equality conditions in the entropy power
inequality; a reverse entropy 
power inequality for log-concave random vectors is proved;
an information-theoretic analog of the
Rogers-Shephard inequality for convex bodies is established;
and it is observed that some of these results lead
to new inequalities for the determinants
of positive-definite matrices. Moreover, by considering
the multiplicative subgroups of the complex plane,
one obtains new inequalities for the differential entropies 
of products and ratios of nonzero, complex-valued
random variables.
\end{abstract}


%


\newpage

\section{Introduction}
\label{sec:intro}

\subsection{Motivation}

\IEEEPARstart{T}{he} properties of the entropy of sums
and differences of random variables have attracted 
a great deal of interest in almost every area of
information theory. Classical results were primarily
motivated by the study of additive noise channels, 
and in the past three decades connections with several
other fields have emerged, including the foundations
of probabilistic limit theorems, functional inequalities
and probabilistic bounds.

More recently, it was also observed that 
inequalities involving the entropies of sums and
differences are closely tied to basic 
questions and results in the area of additive combinatorics, 
which in turn also have applications in communications.
A prominent collection of tools in
additive combinatorics are those provided by the 
Pl\"unnecke-Ruzsa sumset theory;
see, e.g., \cite{TV06:book} for a broad introduction.
A simple example of such a result is the following.
Given two discrete sets $A$ and $B$, the {\em sumset}
$A+B$ is defined as, $A+B=\{a+b\,:\,a\in A,b\in B\}$,
and the {\em difference set} $A-B$ 
is, $A-B=\{a-b\,:\,a\in A,b\in B\}$.
The {\em Ruzsa triangle inequality} \cite{Ruz96:1}
states that, 
for any three sets $A,B,C$, we have,
\be
|A-C|\cdot|B|\leq|A-B|\cdot|B-C|,
\label{eq:triangle}
\ee
where $|E|$ denotes the cardinality of a set $E$,
and $A,B$ and $C$ are subsets of 
the integers, or any other discrete abelian group.
A fascinating connection between such inequalities
and corresponding results for the 
Shannon entropy $H$ was identified initially
by Tao and Vu \cite{TV06:unpub} and by
Ruzsa \cite{Ruz09}, and it has been developed
quite extensively by several authors over the
past 10 years; see, e.g., \cite{MMT12, Tao10}
and the references therein. The main idea is
that, interpreting the entropy as the effective
log-cardinality of the support of a random variable,
then replacing the log-cardinality of every sumset
(or difference set) by the entropy of a corresponding
sum (respectively, difference) of independent discrete
random variables, produces a candidate entropy inequality.
For example, (\ref{eq:triangle}) becomes,
\be
H(X-Z)+H(Y)\leq H(X-Y) + H(Y-Z),
\label{eq:triangleH}
\ee
for independent $X,Y,Z$,
where $H$ denotes the Shannon entropy.

For discrete random variables, this connection was 
studied in detail by \cite{MMT12} and Tao \cite{Tao10}, who established
numerous such entropy inequalities. The main technical
tool in Tao's proofs was the submodularity property
of the discrete entropy, which, as observed in our
subsequent work \cite{KM14}, fails to hold in the case
of differential entropy. Therefore, in order to extend
Tao's results to continuous random variables, new arguments were
necessary, and the key property which replaced submodularity 
in the proofs of almost all of the corresponding 
differential entropy inequalities in \cite{KM14},
was the data processing inequality for mutual information.
As for the results of \cite{MMT12}, some of them can be extended
without too much effort to continuous random variables, while
others rely too much on the bijection-invariance of discrete entropy
and cannot be extended both because of this and because of delicate
measure-theoretic issues that only arise in the continuous case.

The starting point of the present work is the desire
to explore how this family of inequalities can be extended
to random vectors $\RL^n$ and, more generally, to random
variables taking values in general (locally compact, Polish)
abelian groups. Our main results, outlined below,
include unified proofs for many of the
earlier results in \cite{Tao10,MMT12} and all the results of \cite{KM14};
a key ingredient in our approach is the identification of the {\em Ruzsa divergence} as 
the central quantity of interest. 

We note in passing that strong communication-theoretic motivation for the present work
comes from the fact that our results can be used powerfully in the study of 
the degrees of freedom of interference channels (for which the computation of fundamental
limits is a notoriously hard open problem). The results of our prior work \cite{KM14}
played a key role in the works of Wu, Shamai and Verd\'u \cite{WSV15} and Stotz and B\"olcskei \cite{SB14, SB15};
we anticipate that the more general results developed herein will also find applications
to communication theory.

\subsection{Outline of main results}

The first contribution of this work is to 
isolate and study, in a general setting, 
a quantity that plays a key role in the behavior of entropy
of sums and differences; we call this the Ruzsa 
divergence. Let $X$ and $Y$ denote two random variables
which can be discrete, continuous, vector-valued,
or, more generally, taking values in a locally compact
abelian group $G$. The {\em Ruzsa divergence} 
between $X$ and $Y$ is defined as,\footnote{
	Although a symmetrical variant of this quantity, namely $\half (d_R(X\|Y)+d_R(Y\|X))$, has been 
	called the ``Ruzsa distance'' has been studied before 
	in the discrete setting by Tao \cite{Tao10} 
	and for real-valued random variables in \cite{KM14},
	we find that focusing on this non-symmetric version
	makes various developments clearer. 
	Furthermore, the Ruzsa divergence is a particular instance
	of the Kullback-Leibler divergence or relative entropy,
	so that it inherits many of its characteristics,
	but it also has special properties that justify 
	its close study.}
$$d_R(X\| Y):= h(X'-Y')-h(X')=I(X'-Y';Y'),$$ 
where $X'$ and $Y'$ are independent and
have the same marginal distributions as
$X$ and $Y$, respectively, and $h$ denotes
the entropy on $G$.
As described formally in
the following section, $h$ is the usual Shannon
entropy if $G$ is discrete, it is the (joint)
differential entropy when $G=\RL^n$, and
in general it is the entropy defined with 
respect to Haar measure on $G$.
Much of the remainder of this section will summarize how
the basic properties of the Ruzsa divergence can be used
to provide unified proofs for all existing (discrete
and continuous) entropy inequalities in this area, as well
as their extensions to general groups,
offering an analysis on spaces satisfying 
essentially minimal assumptions -- specifically, 
on abelian groups equipped with the minimal 
topological structure necessary to guarantee 
the existence of a Haar measure so that 
a natural notion of entropy can be defined.

The second contribution of this work is to
highlight some interesting connections of
the aforementioned techniques and ideas
with problems related to the
differential entropies of products of positive random variables, 
the entropy power inequality,
results in convex geometry,
and determinantal inequalities.

We begin in Section~\ref{sec:metric} by introducing
the main definitions and assumptions that will remain
in effect throughout the paper. We first formally define
the Ruzsa divergence $d_R(X\|Y)$, as well as two 
related quantities, the conditional Ruzsa divergence and 
the Ruzsa difference. After some elementary observations,
we then state in Theorem~\ref{thm:ruz-tri}
the triangle inequality for $d_R(X\|Y)$, 
which 
implies the inequality~(\ref{eq:triangleH}), and which is seen to be
a simple consequence of a stronger result,
Theorem~\ref{thm:ruz-tri-2}. This is stated
and proved in Section~\ref{sec:pty}, where we also 
establish a number of the important properties of
$d_R(X\|Y)$. In Theorem~\ref{thm:subadd}
we show that it is subadditive with respect
to convolution,
$d_R(X\|Y_1+ Y_2) \leq d_R(X\|Y_1)+ d_R(X\|Y_2),$
and in Theorem~\ref{thm:bsg-full} we give a general
information-theoretic version of the 
Balog-Szemeredi-Gowers theorem,
a significant inequality from additive combinatorics.

In Section~\ref{sec:sumdiff} we first re-interpret
the subadditivity property of Theorem~\ref{thm:subadd}
in the context of important inequalities for the 
cardinalities of sumsets in additive combinatorics,
called the Pl\"unnecke-Ruzsa inequalities.
Specifically, in Theorem~\ref{thm:p-r} we observe that,
if $X,Y_1,Y_2,\ldots,Y_n$ are independent, then,
$$h\left(X-\sum_{i=1}^nY_i\right)
+(n-1)h(X)
\leq
\sum_{i=1}^nh(X-Y_i).
$$
We then examine the 
question of how different the entropies of 
$X+X'$ and $X-X'$ can be, when $X$ and $X'$ 
are independent and identically distributed (i.i.d.).
As was pointed out by Lapidoth and Pete \cite{LP08:ieeei},
the difference between the two can be arbitrarily large, 
which may be rephrased as saying that
$d_R(X\|X')$ and $d_R(X\|-X')$ can differ by an arbitrarily 
large amount. However, in Corollary~\ref{cor:sd-iid}
we show that the ratio between these two Ruzsa 
divergences is always bounded between $1/2$ and 2;
this generalizes the doubling-difference inequality
of \cite{KM14}. In Theorem~\ref{thm:sumdiff}
we give the general version of the 
sum-difference inequality \cite{KM14},
relating $h(X+X')$ and $h(X-X')$
[equivalently, relating $d_R(X\|X')$ and $d_R(X\|-X')$]
when $X$ and $X'$ are independent but
not necessarily
identically distributed.
We close this section by giving general
versions of some recent results
by Wu, Shamai and Verd\'u \cite{WSV15} 
on discrete random variables,
which were used in a study
of the degrees of freedom
of the $M$-user interference channel.
In Lemma~\ref{lem:wt-sums} and
Theorem~\ref{thm:wt-sums} we state
and prove corresponding results
for the entropy of weighted linear 
combinations of random variables
of the form $aX+bY$, where
$X,Y$ take values in a general (locally compact and Polish) abelian
group, and $a,b$ are integers.

In Section~\ref{sec:prod}, we consider the special cases 
of three subgroups $G$ of the complex plane $\mathbb{C}$,
equipped with the multiplication operation: the
half-line $(0,\infty)$, the unit circle $T\subseteq\mathbb{C}$,
and the nonzero complex numbers $\mathbb{C}\setminus\{0\}$.
In each of these cases, the application of our general results
lead to new inequalities for the differential entropies of 
products and ratios of $G$-valued random variables.

In the last four sections we concentrate
on the special case of real random vectors,
taking $G=\RL^n$ and $h$ to be the usual
(joint) differential entropy. In 
Section~\ref{sec:small-d} we look 
at the difference between $h(X+X')$
and $h(X-X')$ from a different perspective,
and provide results in the spirit of
the Freiman-Green-Ruzsa inverse sumset theorems.
In Corollary~\ref{cor:bn} we show
(under certain conditions),
based on a recent result from \cite{BN12},
that if $h(X+X')-2h(X)$ is small, then the distribution
of $X$ is necessarily close to being Gaussian,
in a way that can be precisely quantified in
terms of relative entropy. Then, in Theorem~\ref{thm:gaussC}
we prove a converse result: If the two entropies
$h(X-X')$ and $h(X+X')$ are significantly different,
then the distribution of $X$ will also be significantly
different (in the relative entropy sense) from being
Gaussian. These results can be seen as quantitative
versions of the condition for equality in the
entropy power inequality \cite{Sha48,Sta59}.
Recall that, when applied to i.i.d.\ random vectors
$X,X'$, the entropy power inequality implies that,
$$h(X+X')\geq h(X)+\frac{n}{2}\log 2,$$
where, throughout the paper,
$\log$ denotes the natural logarithm
$\log_e$, so that the entropy and all 
other familiar information-theoretic quantities 
are expressed in nats.
In Section~\ref{sec:repi} we establish
a reverse inequality of this sort:
Corollary~\ref{cor:repi-iid}
states that,
if $X,X'$ are i.i.d.\ with 
a log-concave distribution,
then,
$$h(X+X')\leq h(X)+n\log 2.$$

In Section~\ref{sec:rs} we argue that the
Ruzsa divergence is a natural analog 
of volume-based functionals that arise
in the geometry of convex sets. 
In Corollary~\ref{cor:rs-ent}
we establish the
following information-theoretic analog 
of the Rogers-Shepard inequality: If $X$ 
and $X'$ are i.i.d.\ with
a log-concave distribution on $\RL^n$, 
then,
$$h(X-X')\leq h(X)+2n\log 2.$$
In fact, we conjecture that the
same result holds without the factor
of 2 in the last term above.
Finally, in Section~\ref{sec:det},
we briefly indicate how the earlier 
inequalities for the entropy can be used
to develop corresponding inequalities
for the determinants of positive-definite
matrices. In particular, in Corollary~\ref{cor:det}
we establish the following variant 
of an inequality due to Rotfel'd \cite{Rot67}:
If $K,K_1,K_2,\ldots,K_n$ are positive-definite
matrices,
then,
\ben
\det(K +K_{1}+\ldots+K_{n}) \leq [\det(K)]^{-(n-1)} 
\prod_{j=1}^n \det(K+ K_{j}) .
\een



\newpage

\section{The Ruzsa divergence}
\label{sec:metric}

We begin by introducing the basic definitions of Haar measure
and random variables with values in an abelian group.
Readers not interested in the general formulation can simply
skip to the two main examples below, and 
read the rest of this paper keeping only 
these two key examples in mind.

Let $G$ be an abelian topological group, i.e., a topological space endowed with a
commutative, associative and continuous operation
(i.e., a continuous function from $G\times G$ to $G$ that takes $(x,y)$ to an element of $G$ denoted $x+y$),
which has an identity element 0 (such that $x+0=x$ for all $x$ in $G$)
and with every element having an inverse (i.e., for each $x\in G$
there is an element in $G$ denoted $-x$ such that $x+(-x)=0$).
We will always assume that the topology on $G$ is Polish
(i.e., it is metrizable so that the resulting metric space is 
complete and separable),
and locally compact
(i.e., every point has a compact neighborhood).
The Borel $\sigma$-algebra $\calG$ on $G$ is the $\sigma$-algebra
generated by all open sets.
It is a classical fact (see, e.g., \cite{Hal50:book, Nac76:book, DS14:book}) that
under these assumptions, there exists a (countably additive) measure $\lambda$ defined
on $\calG$ that is translation-invariant, i.e.,  such that
$\lam(A+x)=\lam(A)$ for each $A\in\calG$ and each $x\in G$, where
$A+x=\{a+x:a\in A\}$. Such a measure is called a Haar measure, 
and it is unique
up to scaling by a positive constant. In any given situation, we will assume that
the scaling is chosen at the beginning and fixed; thus we will talk without
further comment about ``the'' Haar measure on $G$.

For our analysis, the normalization (particular scaling chosen) of the Haar measure does not matter.
Nonetheless, it is useful to keep in mind the common normalizations used for the most important examples --
namely discrete groups and the additive group $\RL^n$.
When $G$ is a countable group with the discrete topology, 
we will always take the Haar 
measure $\lambda$ to be counting measure, i.e., 
$\lam(\{g\})=1$ for every element $g\in G$, and 
define $\lambda$ on any subset of $G$ as its 
(possibly infinite) cardinality.
When $G$ is not compact, the Haar measure is infinite, and then it is 
common to fix the normalization by fixing the measure of 
some special set; in the case of $\RL^n$, as usual, by 
requiring $\lam([0,1]^n)=1$,
we obtain the Lebesgue measure.

Let $(\Omega,\calF,\bbP)$ be a probability space, and $X$ be a $G$-valued random variable
on it (i.e., a function from $\Omega$ to $G$ measurable with respect to $\calF$ and $\calG$).
We say that the random variable $X$ taking values in $G$ has a continuous distribution
if its probability distribution, namely the image measure 
$P_X$ induced by the mapping $X$ on $\calG$,
is absolutely continuous with respect to the Haar measure $\lam$. In this case,
denoting the Radon-Nikodym derivative $\frac{dP_X}{d\lam}(x)$ by $f(x)=f_X(x)$,
we say that $X$ has density $f$, and write $X\sim f$. 

\begin{ex}\label{ex:1}
When $G$ is countable, every $G$-valued random variable $X$
has a continuous distribution, and its density is simply the probability mass function of $X$,
i.e., $f_X(x)=\bbP\{X=x\}$.
\end{ex}

\begin{ex}\label{ex:2}
Let $G$ be the set $\RL^n$ equipped with the addition operation, 
so that $\lam$ is the usual Lebesgue measure.
Let $X$ be a $G$-valued random variable. If $X$ is a continuous
random variable, then its density is the usual probability density
function $f_X:G\ra[0,\infty)$ of the random vector $X$
with respect to Lebesgue measure,
satisfying,
\ben
\bbP\{X\in B\}=\int_B f_X(x) dx ,
\een
for each $B\in\calG$, where $\calG$ is the collection of 
Borel subsets of $\RL^n$.
\end{ex}

If $X$ has density $f$ on the group $G$,
the {\it entropy} of $X$ is defined by,
$$
h(X) = - \int_{G} f(x) \log f(x)\,dx,
$$
provided that the integral exists in the Lebesgue sense.
As usual,
we write $h(X)$ even though the entropy depends
only on the density $f$ of $X$. Clearly, $h$ is precisely the discrete entropy
in the setting of Example~\ref{ex:1}, and the differential entropy in the setting of Example~\ref{ex:2}.

To summarize, {\em we assume throughout that $G$ is a 
Polish, locally compact, abelian group},
equipped with the Haar measure $\lambda$ on its
Borel $\sigma$-field ${\cal G}$. Then
it is easy to check that the same properties are
satisfied by the Cartesian product $G^n$ (with 
coordinate-wise addition defining the group structure, 
the product topology defining the topological structure,
and the product measure $\lambda^n$ being its Haar measure), 
for any $n\in\Nat$.
Thus we can define the entropy of any finite collection of 
jointly distributed random variables $(X_1, \ldots, X_n)$,
each with values in $G$, simply by treating $(X_1, \ldots, X_n)$
as a measurable function from $\Omega$ to the Cartesian product $G^n$,
and computing the entropy of its density. [Generally we will not use 
the common term ``joint entropy,'' since we prefer to think of the 
collection of random variables as a single random object.]
In particular, we can define the conditional entropy
between two $G$-valued random elements $X$ and $Y$ by
the usual chain rule, as $h(Y|X)=h(X,Y)-h(X)$.

Although particular care is needed to see which of
the standard properties of discrete entropy and differential
entropy carry over to the general case, we note that it is
immediate from the definition that some key properties remain true.
First, the entropy is always
translation-invariant in that,
for any constant $a\in G$, $h(X+a)=h(X)$, because of the 
translation-invariance
of the Haar measure. Also, the chain rule holds in general,
and, if we define the mutual information as usual as a difference
of entropies, the chain rule for mutual information also
holds in this general setting. Finally, the property which will
play the most central role in our subsequent development,
namely the data processing inequality for mutual information,
also holds in complete generality.

\begin{defn}
Suppose $X$ and $Y$ are $G$-valued random variables with finite entropy.
The {\em Ruzsa divergence} between $X$ and $Y$ is
defined as,
\ben
d_R(X\| Y):= h(X'-Y')-h(X') ,
\een
where $X'$ and $Y'$ are taken to be independent random variables
with the same distributions as $X$ and $Y$, respectively.
\end{defn}

Let us note that even though the entropies of $X$ and $Y$ above are
assumed to be finite, it is possible that $h(X'-Y')$ and hence $d_R(X\| Y)$
are $+\infty$ (see, e.g., \cite{BC15} for examples).
In order to avoid uninteresting technicalities, in  the statements of 
all subsequent definitions and results, {\em we will always implicitly
assume that the entropies and Ruzsa divergences that appear are well-defined
and finite}. The adjustments that need to be made to address possible infinities are left to the reader; 
see, e.g., the discussion after Lemma~\ref{lem:lb} where
we work out explicitly the precise finiteness conditions in one particular case.

A more precise way of writing the Ruzsa divergence
would have been to write it as $d_R(f_1\|f_2)$, where 
$X\sim f_1$ and $Y\sim f_2$, 
but we find it convenient to highlight the random vectors
in the notation. The term ``divergence'' is designed to invoke comparison with
the relative entropy or Kullback-Leibler divergence 
(in that $d_R$ also satisfies some
properties of a distance but not others, e.g., it is
not symmetric); in fact, it is immediately 
obvious
that the Ruzsa divergence is just a special case of the mutual information
(and hence of the relative entropy).

\begin{lem}\label{lem:ruz-mut}
For any two $G$-valued random variables $X,Y$,
\ben
d_R(X\| Y)=I(X'-Y';Y'), 
\een
where $I(Z;W)=h(Z)+h(W)-h(Z,W)$ denotes the mutual information 
between $Z$ and $W$, 
and $X'\sim X$ and 
$Y'\sim Y$ are independent.
In particular, $d_R(X, Y)\geq 0$.
\end{lem}

Observe that $d_R(X\|X)=I(X-X';X)$, where $X'$ is an independent copy of $X$,
and this is rarely identically zero.
In particular, when $G=\RL^n$, $d_R(X\|X)$ is never zero,
since the entropy power inequality
implies a strictly positive lower bound on $d_R(X\|X)$ depending only on $n$,
as discussed in Section~\ref{sec:small-d}.
Thus even if we ignore the assymmetry of Ruzsa divergence (which can be fixed
by averaging $d_R(X\| Y)$ and $d_R(Y\| X)$),  
one should be careful in interpreting it as a notion of distance. 

However, the quantity $d_R$ satisfies a triangle inequality.

\begin{thm}[{\sc Triangle inequality for Ruzsa divergence}]
\label{thm:ruz-tri}
If $X_1,X_2,X_3$ are independent, then,
\ben
d_R(X_1\| X_3)\leq d_R(X_1\| X_2) + d_R(X_2\| X_3) .
\een
\end{thm}


Theorem~\ref{thm:ruz-tri} was proved originally (in an equivalent form) 
for discrete random variables by Ruzsa \cite{Ruz09};
see also Tao \cite{Tao10}.
Since the discrete arguments used in these proofs rely on the property
of submodularity which fails in the continuous setting,
a different proof for Theorem~\ref{thm:ruz-tri} was recently 
provided in \cite{KM14} for real-valued random variables. 
The proof we present for the general
setting in Section~\ref{sec:pty} uses both a re-interpretation 
of the approach used in \cite{KM14},
and 
a sufficient condition for bijections in locally 
compact abelian groups to preserve the entropy,
recently obtained in \cite{MS15} and stated in Lemma~\ref{lem:ent-pres}.


We now define a conditional version of the Ruzsa divergence.
Throughout this paper, we say that $X\leftrightarrow Z\leftrightarrow Y$ 
form a Markov chain if they
are defined on a common probability space and
the conditional distribution of $X$ given $(Z,Y)$ is the same as that of $X$ given $Z$ alone.
The assertion that $X\leftrightarrow Z\leftrightarrow Y$ 
form a Markov chain is easily seen to be symmetric, i.e.,
it is equivalent to the statement that 
$Y\leftrightarrow Z\leftrightarrow X$ form a Markov chain.

\begin{defn}
Suppose  $X_1$, $Y$, and $X_2$ are $G$-valued random variables,
such that $X_1\leftrightarrow Y \leftrightarrow X_2$ 
forms a Markov chain. 
The {\em conditional Ruzsa divergence between $X_1$ and $X_2$ given $Y$}
is, 
\ben
d_R(X_1\| X_2 | Y):= h(X_1-X_2|Y)-h(X_1|Y).
\een
\end{defn}

\begin{lem}\label{lem:cruz-mut}
If $X_1\leftrightarrow Y \leftrightarrow X_2$ form a Markov chain, then,
\ben
d_R(X_1\| X_2 |Y)=I(X_1-X_2;X_2|Y),
\een
where $I(Z;W|V)=h(Z|V)+h(W|V)-h(Z,W|V)$ denotes the 
conditional mutual information 
between $Z$ and $W$, given $V$.
In particular, $d_R(X_1\| X_2 |Y) \geq 0$.
\end{lem}

\begin{IEEEproof}
Observe that,
\ben
d_R(X_1\| X_2 | Y) &=& h(X_1-X_2|Y)-h(X_1|Y) \\
&=& h(X_1-X_2|Y)-h(X_1|Y, X_2) \\
&=& h(X_1-X_2|Y)-h(X_1-X_2 |Y, X_2)\\
&=& I(X_1-X_2;X_2|Y) \\
&\geq& 0 .
\een
The Markov condition was used in an essential way in the second equality of the above display,
while the translation-invariance of entropy was used in the third equality.
\end{IEEEproof}
\vspace{.1in}

Observe that $d_R(X_1\| X_2 |Y)\neq d_R(X_2\| X_1 |Y)$ in general,
but that both quantities are non-negative.

Finally we introduce a more general 
version of the Ruzsa divergence, involving 
dependent random variables.

\begin{defn}
The {\em Ruzsa difference} of the two $G$-valued
random variables $X$ and $Y$ is,
\ben
\tilde{d}_R(X\| Y):= h(X-Y)-h(X) .
\een
\end{defn}

Clearly, $\tilde{d}_R(X\| Y)=d_R(X\| Y)$ when $X$ and $Y$ are independent,
but in general $\tilde{d}_R(X\| Y)$ is not a divergence and need not 
be non-negative.
Indeed, it is easy to see that one always has the following identity.

\begin{lem}\label{lem:ruz-diff}
For any pair of $X,Y$,
\ben
\tilde{d}_R(X\| Y)=I(X-Y;Y) - I(X;Y).
\een
\end{lem}


\newpage

\section{Properties of Ruzsa divergence}
\label{sec:pty}

A special case of the Markov chain condition 
$X_1\leftrightarrow Y \leftrightarrow X_2$ is when
$X_1$ is independent of $(Y,X_2)$. Then, the conditional Ruzsa 
divergence can be related to the
(unconditional) Ruzsa divergence. 

\begin{lem}\label{lem:cond-red}
{\em ({\sc Conditioning reduces Ruzsa divergence})}
If $X_1$ is independent of $(Y,X_2)$, then,
\ben
d_R(X_1\|X_2)= d_R(X_1\| X_2 | Y) + I(Y;X_1-X_2),
\een
and, in particular, $d_R(X_1\| X_2 | Y) \leq d_R(X_1\| X_2)$. 
\end{lem}

\begin{IEEEproof}
By Lemma~\ref{lem:cruz-mut} and the chain rule
for mutual information, 
\ben
d_R(X_1\| X_2 | Y) &=& I(X_1-X_2;X_2|Y) \\
&=& I(X_1-X_2;(X_2,Y)) - I(X_1-X_2;Y) .
\een
But,
\ben
I(X_1-X_2;(X_2,Y))
&=& h(X_1-X_2) - h(X_1-X_2|X_2,Y) \\
&=& h(X_1-X_2) - h(X_1|X_2,Y) ,
\een
by translation-invariance of entropy. 
The assumed 
independence now implies that,
\ben
I(X_1-X_2;(X_2,Y)) &=& h(X_1-X_2) - h(X_1)\\
&=&d_R(X_1\|X_2),
\een
so that,
\ben
d_R(X_1\| X_2 | Y) &=& d_R(X_1\|X_2) - I(Y;X_1-X_2) \\
&\leq& d_R(X_1\|X_2) .
\een
\end{IEEEproof}
\vspace{.1in}

To motivate the next property of Ruzsa divergence 
we will develop, it is useful
to consider the special case $G=\RL^n$, 
equipped with Lebesgue measure. In this case,
it is an elementary fact that for any matrix $A\in GL_n(\RL)$ 
(i.e., any invertible $n\times n$ matrix),
and for any random vector $X$ taking values in $\RL^n$,
$h(AX)=h(X)+\log\det(A)$,
where $\det(\cdot)$ denotes the determinant.
This has two useful consequences. Firstly, 
$d_R(X\|AY) 
= d_R (A^{-1}X\|Y)$ 
so that in particular, $d_R(X\|-Y)=d_R(-X\|Y)$.
Secondly, for any matrix $A\in SL_n(\RL)$ (i.e., 
any invertible matrix with determinant 1),
entropy is preserved by the corresponding linear transformation, i.e.,
$h(AX)=h(X)$.

For a general locally compact abelian group $G$, 
the notion of a linear transformation on $G^n$ defined
by a matrix $A$ no longer makes sense. However, when the 
elements of an $n\times n$ matrix $A=(a_{ij})_{1\leq i,j\leq n}$ 
are integers, we can talk about the group homomorphism 
induced by $A$ on $G^n$. Specifically, for $(x_1, \ldots, x_n) =x\in G^n$, 
we denote by $Ax$ the element,
\ben
\bigg( \sum_{j=1}^n a_{1j}x_j, \sum_{j=1}^n a_{2j}x_j, 
\ldots, \sum_{j=1}^n a_{nj}x_j \bigg)\in G^n,
\een
where $ax$ denotes the element $x+\cdots+x\in G$,
added $a$ times.
Even though $G^n$ is not a linear space, we will sometimes 
call an integer matrix $A$ a ``linear transformation,''
with the understanding that this refers to the group 
homomorphism induced by it as above.

The general linear group over the 
integer ring $\ZZ$ (strictly speaking, of the module $\ZZ^n$), denoted $GL_n(\ZZ)$,
is the set of all $n\times n$ matrices with integer entries 
and determinant $+1$ or $-1$. The following result was 
recently shown in \cite{MS15}.

\begin{lem}\label{lem:ent-pres}
Let $X$ be a random variable taking values in $G^n$.
If $A\in GL_n(\ZZ)$, 
then,
\ben
h(AX)=h(X).
\een
\end{lem}

This allows us to extend the observation that
the Ruzsa divergence behaves nicely when the random vectors involved
are linearly transformed.

\begin{cor}\label{cor:ent-pres}
For any $A\in GL_n(\mathbb{Z})$, and any pair
of $G$-valued random variables $X,Y$,
\ben
d_R(X\|AY) = d_R (A^{-1}X\|Y). 
\een
In particular, $d_R(X\|-Y) = d_R (-X\|Y)$. 
\end{cor}

\begin{IEEEproof}
Assume, without loss of generality, that
$X,Y$ are independent.
By Lemma~\ref{lem:ent-pres},
\ben
d_R(X\|AY) 
&=& h(X-AY)-h(X)\\
&=& h(A^{-1}X-Y) - h(A^{-1}X)\\
&=& d_R (A^{-1}X\|Y) .
\een
\end{IEEEproof}
\vspace{.1in}

We now prove a sharpened version 
of the triangle inequality 
in Theorem~\ref{thm:ruz-tri}.

\begin{thm}\label{thm:ruz-tri-2}
If $X_1,X_2,X_3$ are independent, then,
\ben
d_R(X_1\| X_3)\leq d_R(X_1\| X_2| X_2-X_3) + d_R(X_2\| X_3) .
\een
\end{thm}

\begin{IEEEproof}
By an application of Lemma~\ref{lem:ruz-mut} 
and the data processing inequality
for mutual information, 
\ben
d_R(X_1\| X_3)&=& I(X_1-X_3;X_3)\\
&\leq& I((X_1-X_2, X_2-X_3) ;X_3) .
\een
By the chain rule for mutual information, however,
\ben
&& I((X_1-X_2, X_2-X_3) ;X_3) \\
&=& I(X_2-X_3 ;X_3) + I(X_1-X_2 ;X_3| X_2-X_3) \\ 
&=& d_R(X_2\| X_3) + I(X_1-X_2 ;X_3| X_2-X_3) ,
\een
where we used Lemma~\ref{lem:ruz-mut} 
in the last equality.
All that remains is to show that, 
\ben
d_R(X_1\| X_2| X_2-X_3) 
=I(X_1-X_2 ;X_3| X_2-X_3) ,
\een
or, in view of Lemma~\ref{lem:cruz-mut}, that,
\be\label{eq:trimid}
I(X_1-X_2 ;X_2| X_2-X_3) 
=I(X_1-X_2 ;X_3| X_2-X_3) .
\ee

Let us observe the following general fact:
\be\label{eq:pres-info}
I(X;Y, Y-Z)=I(X;Y,Z) .
\ee
To see this, write,
\ben
I(X;Y, Y-Z)&=&h(Y,Y-Z)-h(Y,Y-Z|X) \\
&=&h(Y,Z)-h(Y,Z|X)\\
&=&I(X;Y,Z) ,
\een
where the second identity relied on Lemma~\ref{lem:ent-pres},
and the fact that the mapping of $(y,z)$ to $(y,y-z)$ is represented by the $2 \times 2$ matrix,
\ben
\left( \begin{array}{cc}
1 & 0 \\
1 & -1
\end{array} \right) ,
\een
which has determinant 1.

Then \eqref{eq:pres-info} implies that,
\ben
I(X_1-X_2 ; X_2, X_2-X_3) 
=I(X_1-X_2 ;X_3, X_2-X_3) ,
\een
since both these quantities equal
$I(X_1-X_2 ;X_2, X_3)$.
Subtracting $I(X_1-X_2;X_2-X_3)$ from both sides,
we obtain the inequality in \eqref{eq:trimid}, completing the proof.
\end{IEEEproof}
\vspace{.1in}

\begin{rmk}
Using Lemma~\ref{lem:cond-red}, Theorem~\ref{thm:ruz-tri-2} 
can be written in a symmetric form as,
\ben
d_R(X_1\| X_3)&\leq& d_R(X_1\| X_2)  + d_R(X_2\| X_3) 
- I(X_1-X_2;X_2-X_3),
\een
and Theorem~\ref{thm:ruz-tri} immediately follows.
\end{rmk}

A useful property of Ruzsa divergence is subadditivity
in the second argument, which
may be equivalently expressed as a monotonicity property 
in the first argument. 

\begin{thm}\label{thm:subadd}
If $X, Y_1$ and $Y_2$ are independent, then,
\ben
d_R(X\|Y_1+ Y_2) \leq d_R(X\|Y_1)+ d_R(X\|Y_2).
\een
Equivalently, if $X_1, X_2$ and $Y$ are independent, then,
\ben
d_R(X_1 + X_2\|Y) \leq d_R(X_1\|Y) .
\een
\end{thm}

\begin{IEEEproof}
Observe that,
\begin{align*}
&d_R(X\|Y_1+ Y_2) - d_R(X\|Y_1) \\
&=h(X-Y_1- Y_2) - h(X)
- [h(X-Y_1)- h(X)] \\
&=h(X-Y_1- Y_2)- h(X-Y_1) \\
&=d_R(X-Y_1;Y_2).
\end{align*}
By relabeling variables, we see that the two formulations 
are equivalent.

To prove the second formulation (and hence also the first), note that
by Lemma~\ref{lem:ruz-mut}, and the 
data processing inequality and the chain rule for mutual information, 
\ben
d_R(X_1 + X_2\|Y) &=& I(X_1 + X_2-Y;Y)\\
&\leq& I(X_1 -Y, X_2;Y)\\
&=& I(X_1 -Y;Y) + I(X_2;Y| X_1 -Y) .
\een
The second term in the last line is 0 since $X_2$ is independent of $(X_1,Y)$, 
so that another application of Lemma~\ref{lem:ruz-mut} gives the desired result.
\end{IEEEproof}
\vspace{.1in}

\begin{rmk}\label{rmk:plu-r-hist}
Written out in terms of entropies, Theorem~\ref{thm:subadd}
is equivalent to the assertion that the entropy of a sum of independent group-valued random variables
is a submodular set function, i.e.,
$h(X+Y+Z)+h(Z) \leq h(X+Z)+h(Y+Z)$.
For discrete entropy, this assertion is implicit
in Kaimanovich and Vershik \cite{KV83}, and explicitly and independently developed
in \cite{Mad08:itw, MMT12}; \cite{MMT12} also contains a generalization from sums
to a more general class of so-called partition-determined functions that can make sense
on sets with less algebraic structure.
For differential entropy, this assertion was first presented in \cite{Mad08:itw},
and further explored for the case of $\RL$-valued random variables in \cite{KM14}.
\end{rmk}

If we do not make assumptions about the nature of the underlying distributions,
the Ruzsa divergence and conditional Ruzsa divergence can be unbounded.
In Sections~\ref{sec:repi} and \ref{sec:rs}, 
we will make such assumptions and
demonstrate a uniform bound on Ruzsa divergence for a 
log-concave density on $\RL^n$.
On the other hand, it is possible to obtain a bound on conditional Ruzsa divergence
under mild assumptions on the dependence structure.

\begin{thm}\label{thm:cond-ruz-bd}
If $X_1\leftrightarrow Y \leftrightarrow X_2$ form a Markov chain, then,
\ben
d_R(X_1\|X_2|Y) &\leq  2 I(X_1;Y) + I(X_2;Y) +
 \tilde{d}_R(X_1\|Y) + \tilde{d}_R(Y\|X_2) .
\een
\end{thm}

\begin{IEEEproof}
Let $(X_1,Y,X_2)$ and 
$(X_1,Y'',X_2)$ be conditionally independent
versions of 
$(X_1,Y,X_2)$, given $(X_1,X_2)$.
By the data processing inequality:
\begin{align*}
& I(X_1-X_2; X_1|Y) \\
&\leq I(X_1+Y'', X_2+Y''; X_1|Y) \\
&= h(X_1|Y)
	+h(X_1+Y'', X_2+Y''|Y) \\
	&\quad -h(X_1+Y'', X_2+Y'', X_1|Y)\\
&= h(X_1|Y)
	+h(X_1+Y'', X_2+Y''|Y)
	-h(X_1,X_2,Y''|Y),
\end{align*}
where the last equality follows from Lemma~\ref{lem:ent-pres}, and the fact
that the linear map $(x_1,x_2,y)\mapsto(x_1+y,x_2+y,x_1)$
has determinant~$-1$. Therefore,
\begin{align*} 
h(X_1-X_2|Y)
&= h(X_1-X_2|X_1,Y) + I(X_1-X_2; X_1|Y)\\
&= h(X_2|Y) +  I(X_1-X_2; X_1|Y)\\
&\leq  h(X_1,X_2|Y)
	+h(X_1+Y'', X_2+Y''|Y) \\
	&\quad -h(X_1,X_2,Y''|Y) .
\end{align*}
We have established that,
\be\label{eq:Csub}
h( X_1,X_2,Y,Y'' ) + h(X_1-X_2, Y)
\leq h(X_1,X_2,Y) + h(X_1+Y'', X_2+Y'',Y) .
\ee

We now deduce the result from~(\ref{eq:Csub}).
First note that by conditional independence of $Y$ and $Y''$ given $X_1, X_2$,
the first term in the left-hand of (\ref{eq:Csub}) is,
$$h( X_1,X_2,Y,Y'' ) +h(X_1, X_2)
= h(X_1,X_2,Y)+h(X_1,X_2, Y'') 
= 2 h(X_1,X_2, Y),$$
so that,
\begin{align*} 
h(X_1-X_2, Y) 
&\leq  h(X_1+Y'', X_2+Y'',Y) \\
        &\quad - h(X_1,X_2,Y) +h(X_1,X_2)\\
&\leq \sum_i h(X_i+Y) + h(Y) \\
        &\quad - h(X_1,X_2,Y) +h(X_1,X_2).
\end{align*}

By conditional independence and the chain rule,
\begin{align*} 
h(X_1,X_2,Y)
&= h(X_1,X_2|Y)+h(Y)\\
&= h(X_1|Y)+ h(X_2|Y)+h(Y) .
\end{align*}
Thus,
\ben
h(X_1-X_2|Y) + h(Y)
&\leq& \sum_i h(X_i+Y) + h(Y) +h(X_1,X_2) \\
&&-[h(X_1|Y)+ h(X_2|Y)+h(Y) ] \\
&=& h(X_1+Y) - h(X_1|Y)  \\
&&+ h(X_2+Y) - h(X_2|Y) +h(X_1,X_2).
\een
So,
\begin{align*}
&h(X_1-X_2|Y) - h(X_1|Y) \\
&\leq I(X_1;Y) + \tilde{d}_R(X_1\|Y) 
+ h(X_2+Y) - h(X_2|Y) +h(X_1,X_2)- h(X_1,Y).
\end{align*}
Since,
\begin{align*}
h(X_2|Y) -h(X_1,X_2)+ h(X_1,Y)
&= h(X_1, X_2|Y)+h(Y)-h(X_1,X_2)\\
&= h(Y|X_1, X_2) \\
&= h(Y|X_2) -I(Y;X_1| X_2),
\end{align*}
and since,
\begin{align*} 
I(Y;X_1| X_2)&= h(X_1|X_2)-h(X_1|Y, X_2)\\
&= h(X_1|X_2)-h(X_1|Y) \\
&\leq h(X_1)-h(X_1|Y) \\
&= I(X_1;Y) , 
\end{align*}
we are done.
%
\end{IEEEproof}
\vspace{.1in}

Let us note two corollaries of Theorem~\ref{thm:cond-ruz-bd}.
Firstly, if we assume $X_1, X_2$ and $Y$ to be independent,
we recover the Ruzsa triangle inequality (Theorem~\ref{thm:ruz-tri}).
Secondly, the case where the joint distribution is symmetric in $(X_1, X_2)$
is of interest.

\begin{cor}\label{cor:cond-ruz-bd}
Suppose $X_1\leftrightarrow Y \leftrightarrow X_2$ form a Markov chain,
and $X_1$ and $X_2$ have the same conditional distribution given $Y$. 
Then,
\ben
d_R(X_1\|X_2|Y) \leq  3 I(X;Y) +\tilde{d}_R(X\|Y) + \tilde{d}_R(Y\|X) .
\een
\end{cor}

One may interpret this as follows. For every possible value $y$ of $Y$, 
consider the Ruzsa divergence between the conditional distribution of $X$ given $Y=y$, and itself;
then the conditional Ruzsa divergence $d_R(X_1\|X_2|Y)$ is the average of these quantities under
the distribution of $Y$. This follows from the fact that $X_1, X_2$ are conditionally i.i.d. given $Y$.
Thus Corollary~\ref{cor:cond-ruz-bd} says that,
for weakly dependent random variables $X, Y$,
having bounds on the two (not particularly well behaved) Ruzsa 
differences between $X$ and $Y$, allows one 
to get a bound on this averaged self-divergence of the conditional distribution 
of $X$ given $Y$ (which is a well behaved divergence).



Let us recall the Balog-Szemeredi-Gowers theorem, which has become an extremely
useful tool in additive combinatorics in the last two decades.
There are several formulations, but the one we focus on is 
stated in terms of the {\em restricted sumset} $A\plusE B$,
defined as,
$$
A\plusE B
=\{a+b:a\in A, b\in B, (a,b)\in E\},$$
where $E$ is some subset of the Cartesian product $A\times B$.
If $A$ and $B$ are finite nonempty subsets of an abelian group $G$, 
and $E\subset A\times B$ satisfies $|E|\geq \frac{1}{K} |A|\cdot|B|$
and $|A\plusE B| \leq K\sqrt{|A|\cdot|B|}$ for some $K\geq 1$,
then there exist subsets $A_0\subset A$ and $B_0\subset B$ such that
$|A_0|\geq \frac{1}{K}|A|$, $|B_0|\geq \frac{1}{K}|B|$, and
\ben
|A_0 + B_0 | \leq K^7 \sqrt{|A_0|\cdot|B_0|} .
\een
The natural probabilistic analogue of a restricted 
sumset is a sum of dependent random variables.
Theorem~\ref{thm:cond-ruz-bd} may be thought of as an 
information-theoretic form of the 
Balog-Szemeredi-Gowers theorem, since bounds for dependent random vectors
are used to deduce bounds for (conditionally) independent  random vectors.
It is not directly analogous to the Balog-Szemeredi-Gowers theorem
since the bounds are not in terms of the Ruzsa differences between $X_1$ and $X_2$,
but rather in terms of the Ruzsa differences between either of them and the auxiliary random variable $Y$.
However, such a direct analogue can be constructed 
using Theorem~\ref{thm:cond-ruz-bd}.
This was done in the discrete case by Tao \cite{Tao10}, and in the case of the additive group $\RL$ by the authors in \cite{KM14}.
We state below the resulting theorem in the general setting, 
using the notation developed in this paper.

\begin{thm}\label{thm:bsg-full}
Let $(X_2,Y_1,X_1,Y_2)$ form a Markov chain,
with the marginal distributions of the 
pairs $(X_2, Y_1), (X_1, Y_1)$ and $(X_1, Y_2)$
all being the same as the distribution of $(X,Y)$.
Then,
\ben
d_R(X_2\|Y_2 | X_1, Y_1) + d_R(Y_2\|X_2 | X_1, Y_1) 
\leq 3 I(X;Y) + \tilde{d}_R(X\|Y)+\tilde{d}_R(Y\|X) .
\een
\end{thm}

\vspace{.1in}\begin{IEEEproof}
The proof of  \cite[Theorem 3.14]{KM14} for real-valued random 
variables carries over
almost exactly in the general case, if one uses Lemma~\ref{lem:ent-pres}
to justify one of the steps. This yields,
under the present assumptions,
that,
\ben
I(X_2+Y_2;Y_2|X_1, Y_1) + I(X_2+Y_2;X_2|X_1, Y_1) 
\leq I(X;Y) + I(X+Y;X)+I(X+Y;Y) .
\een
To obtain the desired result in the stated form, one just 
needs to replace all occurrences of $Y, Y_1$ or $Y_2$ by their
respective inverses (i.e., $-Y, -Y_1$ or $-Y_2$), and then make 
appropriate use of
Lemma~\ref{lem:ruz-mut}, Lemma~\ref{lem:cruz-mut}, 
and Lemma~\ref{lem:ent-pres}.
\end{IEEEproof}

\newpage

\section{Entropies of weighted sums and differences}
\label{sec:sumdiff}

The Pl\"unnecke inequality in additive combinatorics
\cite{Plu70, Ruz89, Ruz90} states that, if $|A+ B|\leq \alpha |A|$
for finite nonempty subsets $A, B$ of an abelian group, 
then for every $k>1$, there exists a nonempty subset $A'\subset A$ such that 
\be\label{orig-plu}
|A'+kB|\leq \alpha^{k} |A'| ,
\ee
where $kB$ refers to the sumset $B+\cdots+B$ with $k$ summands. 
A very elegant and considerably simpler proof, obtained by Petridis \cite{Pet12},
also shows that the same subset $A'$ can be used for all positive integers $k$.
The inequality \eqref{orig-plu} can be generalized to different summands:
if $A$ and $B_i$ are nonempty finite sets, with
$|A+B_i|\leq \alpha_i |A|$ for each $i$, then there exists a nonempty subset $A'\subset A$ such that,
\ben
|A'+B_1+\ldots+ B_m|\leq \bigg(\prod_{i=1}^m \alpha_i \bigg) |A'|.
\een
This is usually called the Pl\"unnecke-Ruzsa inequality, since it was proved by Ruzsa \cite{Ruz89, Ruz90}
using an ingenious combinatorial argument. 
These inequalities are very influential in additive combinatorics--
for example, as expounded in \cite{TV06:book}, they are sufficient to obtain Freiman-type
inverse theorems for groups with bounded torsion. 
The analogue of the Pl\"unnecke-Ruzsa inequality for the
entropy is the following subadditivity property of Ruzsa divergence, 
which is an immediate consequence of Theorem~\ref{thm:subadd};
the same historical remarks made in Remark~\ref{rmk:plu-r-hist}
therefore also apply here.

\begin{thm}\label{thm:p-r}
If $X, Y_1, \ldots, Y_k$ are independent, then:
\ben
d_R\bigg(X \bigg\| \sum_{i=1}^k Y_i\bigg) \leq \sum_{i=1}^k d_R(X\|Y_i) .
\een
\end{thm}

To see that this is analogous to the Pl\"unnecke-Ruzsa inequality as stated above,
we can trivially rewrite it
in the following form: if $d_R(X\|Y_i)\leq \alpha_i$, then
$d_R(X\| \sum_{i=1}^k Y_i)\leq \sum_{i=1}^k \alpha_i$.
Unlike in the case of sets where one potentially needs to pass to a subset to obtain a valid inequality,
the entropy analogue works with the original random variables of interest.



The properties of Ruzsa divergence developed in Section~\ref{sec:pty} can also be
used to understand how the differential entropy of the sum of two independent random vectors
constrains the differential entropy of their difference.

\begin{thm}\label{thm:sumdiff}
For any $G$-valued random variables $X, Y$,
\ben
d_R(X\| -Y) \leq 2 d_R(X\|Y) + d_R(Y\|X) .
\een
\end{thm}

\begin{IEEEproof}
Let $(X_1,Y_1)$ be independent,
with  $Z=X_1-Y_1$. Assume $(X_2,Y_2)$ is conditionally independent 
of $(X_1,Y_1)$ given $Z$, and has the same conditional distribution
given $Z$ as $(X_1,Y_1)$; thus in particular $Z=X_2-Y_2$.
Let $(X,Y)$ be independent of $(X_1,Y_1,X_2,Y_2)$,
but have the same distribution as either pair $(X_i,Y_i)$.

Since, by construction, $X_1-Y_1=X_2-Y_2=Z$,
\ben
X+Y &=& X+Y+(X_2-Y_2)-(X_1-Y_1)\\
&=& (X-Y_2)-(X_1-Y)+X_2+Y_1,
\een
and hence, by data processing for mutual information,
\ben
I(X;X+Y)
&\leq&  I(X; X-Y_2,X_1-Y,X_2,Y_1) \\
&=& h(X-Y_2,X_1-Y,X_2,Y_1) \\
      &&\quad -h(X-Y_2,X_1-Y,X_2,Y_1 | X)\\
&=&   h(X-Y_2,X_1-Y,X_2,Y_1)
	-h(Z,Y_1,Y_2,Y |X),
\een
where the last equality follows from the fact 
that the linear map,
$(z,y_1,y_2,y,x)\mapsto(x-y_2,y_1+z-y,y_2+z,y_1,x)$,
has determinant~$1$. 

Using the 
independence of $X$ and $Y$ from each other 
and all other random variables for the second term 
on the above right-hand side,
we have,
\begin{align}
d_R(Y\|-X) 
&\leq  h(X-Y_2) + h(X_1-Y) + h(X_2) +h(Y_1) \nonumber\\
        & \quad -[ h(Z,Y_1,Y_2) + h(Y)] \nonumber\\
&= [d_R(Y\|X) +h(Y)] +[d_R(X\|Y) +h(X)] \nonumber\\
&\quad +h(X_2) -h(Z,Y_1,Y_2).
\label{eq:mm1}
\end{align}
However, observe that, since $I(Y_1;Y_2|Z)=0$, 
\begin{align}
h(Z,Y_1,Y_2)+h(Z)&= h(Y_1, Z)+h(Y_2, Z) \nonumber\\
&=h(X_1, Y_1)+h(X_2, Y_2)
= 2h(X,Y).
\label{eq:MIC}
\end{align}
Plugging \eqref{eq:MIC} into \eqref{eq:mm1} gives,
\ben
d_R(Y\|-X) 
&\leq& d_R(Y\|X) +d_R(X\|Y) +h(Y)+2h(X) \\
&&\quad -[2h(X,Y)- h(Z)] \\
&=& d_R(Y\|X) +d_R(X\|Y) +h(Z)-h(Y)\\
&=& 2d_R(Y\|X) +d_R(X\|Y) ,
\een
which is the desired result.
\end{IEEEproof}
\vspace{.1in}

In the case where $X$ and $Y$ are not just independent but also 
identically distributed, Theorem~\ref{thm:sumdiff} simply says that
$d_R(X\| -X) \leq 3 d_R(X\|X)$, while taking $X$ and $-Y$ 
to have the same distribution gives,
$$d_R(X\|X)\leq d_R(X\| -X) + 2 d_R(X\| -X) = 3 d_R(X\| -X).$$
In fact, one can obtain tighter bounds in these special cases.

\begin{cor}\label{cor:sd-iid}
If $X, Y$ are i.i.d., then:
\ben
\frac{d_R(X\| -X)}{d_R(X\| X)} \in [\half, 2] .
\een
\end{cor}

\begin{IEEEproof}
The desired statement is equivalent, for $X_1,X_2$ that are  i.i.d., to:
\be\label{eq:ent-sd}
\frac{1}{2}
\leq
\frac{h(X_1+X_2)-h(X_1)}{h(X_1-X_2)-h(X_1)}
\leq 2.
\ee
As observed in \cite{KM14}, the upper bound in the inequality 
\eqref{eq:ent-sd} follows 
from Theorem~\ref{thm:p-r},
and the lower bound follows from Theorem~\ref{thm:ruz-tri},
both of which we have already proved for the general setting.
\end{IEEEproof}
\vspace{.1cm}

Corollary~\ref{cor:sd-iid} provides inequalities between $h(X+Y)$ and $h(X-Y)$ when
$X, Y$ are i.i.d. and $h(X)$ is known. The requirement to know
$h(X)$ to make the comparison cannot be dispensed with in the general setting
of locally compact abelian groups. However, this requirement can be dispensed with for discrete groups--
as observed by \cite{ALM15}, $h(X+Y)/h(X-Y)$ must lie between 3/4 and 4/3
if $X$ and $Y$ are i.i.d. random variables in a discrete group.

Finally, let us examine what can be said about
weighted sums and differences, i.e., about
random variables of the form $aX+bY$ where $a, b$ are non-zero integers.
Discrete entropy inequalities for such random variables play a key role in the
recent work of Wu, Shamai and Verd\'u \cite{WSV15} on the degrees of freedom
of the $M$-user interference channel -- specifically, they immediately yield inequalities of similar
form for the R\'enyi information dimension of weighted sums of random variables,
which imply, using the single-letter characterization of \cite{WSV15},
that for rational channel coefficients the number of degrees of freedom is strictly smaller than $M/2$.
In the following theorem, we extend all the inequalities proved by \cite{WSV15} for discrete entropy
of weighted sums and differences to the general abelian setting. 
First we give the generalization of \cite[Lemma 18]{WSV15}.

\begin{lem}\label{lem:wt-sums}
Let $X, X'$ and $Z$ be independent $G$-valued random variables,
where $X'$ has the same distribution as $X$. 
Let $a, b$ be nonzero integers. Then:
\ben
h(aX+b) \leq h((a-b)X+bX'+Z) + d_R(X\|X) .
\een
Furthermore, if $a$ is even, then:
\ben
h(aX+b) \leq h\bigg(\frac{a}{2} X+Z\bigg) + h(2X-X') - h(X) .
\een
\end{lem}

\begin{IEEEproof}
One can simply follow the proof strategy of \cite[Lemma 18]{WSV15}, which on inspection relies only
on the subadditivity of Ruzsa divergence and the Ruzsa triangle inequality, both of which we have already proved
in the general setting.
\end{IEEEproof}
\vspace{.1cm}

Finally we give the generalization of \cite[Theorem 14]{WSV15};
its proof is again the same as in the discrete case,
using the subadditivity of Ruzsa divergence 
and Ruzsa triangle inequality established earlier.
The result of Theorem~\ref{thm:wt-sums} can be compared to 
the inequalities of Bukh \cite{Buk08} for dilated sums of sets.

\begin{thm}\label{thm:wt-sums}
Let $X$ and $Y$ be independent $G$-valued random variables, 
and $a, b$ be nonnegative integers. Then,
\ben
h(aX+bY)-h(X+Y) \leq \tau_{a,b} \, 
\big\{ d_R(X\|-Y) + d_R(Y\|-X) \big\} , 
\een
where,
\ben
\tau_{a,b}=6\big(\floor{\log |a|}  + \floor{\log |b|} +2\big) .
\een
\end{thm}


\newpage

\section{Entropies of products and ratios}
\label{sec:prod}

Since we will need to discuss entropies with respect to 
two different measures on the same group, we introduce
some additional notation to keep things unambiguous.
All the examples considered in this section
involve subgroups $G$ of the group 
$\mathbb{C}^{\times}=\mathbb{C}\setminus\{0\}$
equipped with the multiplication operation. 
The Haar measure for such multiplicative groups
is typically not the same as the familiar Lebesgue measure 
used to compute differential entropies of real-valued
or complex-valued random variables 
(the one-dimensional and two-dimensional Lebesgue 
measures are Haar measures 
for the groups $\RL$ and $\mathbb{C}$ respectively, 
but only when the group structure comes from the addition operation).

\subsection{Positive random variables}

Consider the group $\RL_{>0}=(0,\infty)$ equipped with the 
multiplication operation. Its
Haar measure is given by,
\ben
\lam(dx)=\frac{dx}{x} ,
\een
where $dx$ is Lebesgue measure 
on $(0,\infty)$. To see this,
all we need to do is check the translation-invariance of $\lam$ with 
respect to multiplication,
i.e., that for any fixed $c>0$, we have $\lam(cA)=\lam(A)$
when 
\ben
\lam(A)=\int_A \frac{dx}{x} ,
\een
and $dx$ represents Lebesgue measure on $\RL$.
[And this in turn is an immediate consequence
of the fact that the logarithmic function is an isomorphism between
$(\RL_{>0},\times)$ and $(\RL,+)$, using the 
standard translation-invariance
of Lebesgue measure for addition.]

We are interested in two entropies of a positive (i.e., $\RL_{>0}$-valued) random variable $X$. To 
define them, let us assume that $X$ has a density $f$ with respect to Lebesgue measure on $(0,\infty)$.
Then:
\begin{enumerate}
\item The differential entropy of $X$ is,
\ben
h_{\RL}(X)=-\int_0^\infty f(x)\log f(x) dx .
\een
\item The intrinsic entropy $h_{\times}(X)$ with respect to Haar 
measure $\lam$ on $(\RL_{>0},\times)$ is given by,
\be\label{eq:rplus-ents}
h_{\times}(X)= -\int _0^\infty [xf(x)]\log [xf(x)] \lam(dx)
= h_{\RL}(X) -\E[\log X] ,
\ee
since the density of $X$ with respect to $\lam$ is $xf(x)$.
We use $h_{\times}$ to emphasize that this is the intrinsic entropy with respect to the multiplicative structure 
on $\RL_{>0}$ rather than the additive structure on $\RL$.
\end{enumerate}

Observe that $\RL_{>0}$ is a Polish, locally compact, abelian group to which all of our preceding results
apply and yield statements of interest. For illustration, we only write out one consequence:
Corollary~\ref{cor:sd-iid} says that,
\ben
\frac{1}{2}
\leq
\frac{h_{\times}(X Y)-h_{\times}(X)}{h_{\times}(X/Y)-h_{\times}(X)}
\leq 2,
\een
which,
using relation \eqref{eq:rplus-ents},
translates to the following statement for the usual differential entropy.

\begin{cor}\label{cor:rplus}
If $X, Y$ are i.i.d. random variables taking values in $(0,\infty)$, then:
\ben
h_{\RL} (XY) &\leq& 2h_{\RL} (X/Y)- h_{\RL} (X) + 3\E[\log X] ,\\
h_{\RL} (X/Y) &\leq& 2h_{\RL} (XY)- h_{\RL} (X) - 3\E[\log X]  .
\een
\end{cor}

\subsection{Random variables on the circle group}

Consider the unit circle $T=\{z\in\mathbb{C}: |z|=1\}$
in the complex plane;
this is of course a group under multiplication, and is isomorphic to
$\RL/\ZZ$ equipped with addition via the isomorphism $t\mapsto e^{2\pi i t}$. Alternatively 
we can parametrize $T$ using the angle $\theta$ subtended by the arc of the circle
between the point on $T$ and the real axis (which is just $2\pi t$).
With this parametrization, the Haar measure $\lambda$
is the uniform distribution
on the angle or, equivalently, Lebesgue measure on $[0,2\pi)$.
For a $T$-valued random variable $\Theta$ that has a density 
$f$ with respect to 
the uniform measure, 
\ben
h(\Theta)=-\int_T f(x)\log f(x) \lambda(dx) = - D(\Theta\|U) ,
\een
where $U\sim\lambda$ is uniformly distributed on 
$T$, and $D(\Theta\|U)$ denotes the relative between
$\Theta$ and a uniformly distributed random variable
$U$ on $T$. Thus, the fact that entropy
increases on convolution captures in this setting the fact that convolution brings
any distribution closer to the uniform.

In this case, Corollary~\ref{cor:sd-iid} becomes the following statement.

\begin{cor}\label{cor:circle}
If $\Theta, \Theta'$ are i.i.d. random variables taking values in $T$, then:
\ben
\frac{1}{2}
\leq
\frac{D(\Theta\|U)- D(\Theta+\Theta'\|U)}{D(\Theta\|U)- D(\Theta-\Theta'\|U)}
\leq 2 .
\een
\end{cor}

\subsection{Non-zero complex random variables}

Finally we consider the full group $(\mathbb{C}^{\times},\times)$, 
whose Haar measure
is given by,
\ben
\frac{dz}{|z|^2} ,
\een
where $dz$ is 2-dimensional Lebesgue 
measure (using the identification
of $\mathbb{C}$ with $\RL^2$). 
If $f$ is the density of a $\mathbb{C}^{\times}$-valued random variable $Z$
with respect to 2-dimensional Lebesgue measure, one has the intrinsic entropy,
\ben
h_{\times}(Z)= -\int_{\mathbb{C}^{\times}} [|z|^2 f(z)]\log [|z|^2 f(z)] \frac{dz}{|z|^2}
= h_{\RL^2}(Z) -\E[\log (|Z|^2)] ,
\een
where we use $h_{\RL^2}(Z)$ to denote the usual differential entropy of $Z$.

Then Corollary~\ref{cor:sd-iid} becomes the following statement.

\begin{cor}\label{cor:rplus2}
If $Z_1, Z_2$ are i.i.d. random variables taking values in $\mathbb{C}^{\times}$, then:
\ben
h_{\RL^2} (Z_1 Z_2) &\leq& 2h_{\RL^2} (Z_1/Z_2)- h_{\RL^2} (Z_1) + 6\E[\log |Z_1|] ,\\
h_{\RL^2} (Z_1/Z_2) &\leq& 2h_{\RL^2} (Z_1 Z_2)- h_{\RL^2} (Z_1) - 6\E[\log |Z_1|]  .
\een
\end{cor}


\newpage

\section{Freiman-type results for the entropy on $\RL^n$}
\label{sec:small-d}

For the rest of the paper, our focus is on the additive group $\RL^n$
equipped with Lebesgue measure, so that $h$ denotes the usual differential entropy.
Our first observation is a uniform lower bound on the Ruzsa divergence
between a distribution and itself. A simple application of 
the entropy power inequality \cite{Sha48}\cite{Sta59} to two i.i.d.\
random variables easily gives the following result.

\begin{lem}\label{lem:lb}
For any $\RL^n$-valued random vector $X$ with finite differential entropy,
\ben
d_R(X\|X) \geq \frac{n}{2} \log {2} . 
\een
Furthermore, $d_R(X\|-X) \geq \frac{n}{2} \log {2}$. 
\end{lem}

The assumption of finite differential entropy in Lemma~\ref{lem:lb} is in fact essential.
As shown by Bobkov and Chistyakov \cite[Proposition 1]{BC15}, there exists a
$\RL$-valued random variable $X$ of finite entropy such that if $X, X'$ are i.i.d., 
the entropy of $X+X'$ does not exist. 
However, \cite{BC15} also shows that for any such example, necessarily the entropy of $X$ is 
$-\infty$, so that it remains true that if the entropy exists and is a real number, 
then the entropy of the self-convolution also exists (although, thanks to another example
constructed in \cite{BC15}, it may then be $+\infty$!). Henceforth, as stated in Section~\ref{sec:metric},
if nothing is stated, we will assume that all entropies and Ruzsa divergences exist and are finite.


%
%
%

We find it convenient to restate Lemma~\ref{lem:lb} in terms of 
the doubling and difference constants
associated with a random vector.

\begin{defn}
For an $\RL^n$-valued random vector $X$, 
the entropy power of $X$ is defined as,
\ben
\calN(X)=\exp\Big\{\frac{2h(X)}{n}\Big\} .
\een
For an $\RL^n$-valued random vector $X$,
the doubling constant is defined by,
\ben
\sigma_{+}(X)= \frac{\calN(X+X')}{2\calN(X)},
\een
and the difference constant is defined by,
\ben
\sigma_{-}(X)= \frac{\calN(X-X')}{2\calN(X)} ,
\een
where $X'$ is an independent copy of $X$.
\end{defn}

Then entropy power inequality immediately implies that if $X$ has finite entropy, then
$\sigma_{+}(X)\geq 1$ 
and
$\sigma_{-}(X)\geq 1$;
this is just a restatement of Lemma~\ref{lem:lb} since,
\be\label{eq:ruz-diff}
\sigma_{-}(X)=\half \exp\bigg\{\frac{2}{n} d_R(X\|X)\bigg\},
\ee
and,
\be\label{eq:ruz-doub}
\sigma_{+}(X)=\half \exp\bigg\{\frac{2}{n} d_R(X\|-X)\bigg\} .
\ee
Furthermore, because of the equality conditions of the entropy power inequality,
$\sigma_{+}(X)$ (or $\sigma_{-}(X)$) is equal to 1 if and only if
$X$ is a Gaussian (with non-singular covariance matrix).
Note that the definitions of doubling and difference
constants of scalar random variables in \cite{Tao10} 
(for discrete random variables)
and in \cite{KM14} (for $\RL$-valued random variables) used a different
normalization, but we have chosen the normalization above so that the minimum
value achieved at Gaussians for both $\sigma_+$ and $\sigma_-$ is 1.

A natural question is whether the extremality of Gaussians is 
a stable phenomenon.
In other words, if $\sigma_{+}(X)\leq K$ for some $K$, does this
imply that the distribution of $X$ is necessarily not far from
being Gaussian, in a sense that can be quantified in terms of $K$?
It is a perhaps somewhat surprising result due to Bobkov, 
Chistyakov and G\"otze \cite{BCG13:cramer}
that the answer is ``no,'' even in the one-dimensional setting.
Nonetheless, as observed in \cite{KM14},
under the additional assumption that $X$ has a 
finite Poincar\'e constant
(and using results independently obtained by Johnson and Barron \cite{JB04} 
and Artstein, Ball, Barthe and Naor \cite{ABBN04:2}
on the rate of convergence in the information-theoretic central limit 
theorem for 
$\RL$-valued random variables) it can be shown that such a
stability bound can indeed be established.
This result cannot be directly extended to the case 
of $\RL^n$-valued random vectors, since non-asymptotic
bounds that exhibit convergence rates for the entropic central limit theorem 
in the multivariate case are not known under just 
the assumption of a finite Poincar\'e constant\footnote{While asymptotic estimates
of this sort are known \cite{BCG13:rate, BCG14:be, CS11}, estimates that only hold
for a sufficiently large number of summands are not strong enough for our purposes.}. However, 
by relying on recent work of Ball and Nguyen \cite{BN12},
one can see that such stability does hold under the stronger 
assumption of log-concavity. 

Recall that a probability density function 
$f$ defined on 
$\RL^n$ is said to be log-concave if,
\ben
f(\alpha x +(1-\alpha)y) \geq f(x)^{\alpha} f(y)^{1-\alpha} ,
\een
for each $x,y\in \RL^n$ and each $0\leq \alpha\leq 1$. If $f$ is log-concave, we will
also use the adjective ``log-concave'' for a random variable $X$ distributed according
to $f$, and for the probability measure induced by it. 
Note that the class of log-concave probability measures is quite broad,
including the uniform distribution on any compact, convex set, 
the exponential distribution, and of course any Gaussian. 
On the other hand, log-concavity can also be fairly restricting: 
For instance, it implies at least exponentially decaying tails, 
and a finite Poincar\'e constant.

Now we state the main result of  \cite{BN12} we will need.
For a random vector $X\sim f$ we write $D(X)$ for its
relative entropy distance from a Gaussian,
\ben
D(X)=D(f\|f^G)= h(f^G)-h(f) ,
\een
where $f^G$ is the Gaussian density with the same mean 
and covariance matrix as $f$,
and $D$ is the usual relative entropy.

\begin{thm}\label{thm:bn}\cite{BN12}
Suppose $X$ is a log-concave random vector in $\RL^n$, 
and that it satisfies a Poincar\'e inequality with constant $c$, i.e.,
if for any smooth function $u$ with $E[u(X)]=0$,
\ben
cE [u(X)^2] \leq E [|\nabla u(X)|^2] .
\een
Then,
\ben
h\bigg(\frac{X_1+X_2}{\sqrt{2}}\bigg) -h(X) \geq \frac{c}{4(1+c)} D(X) ,
\een 
where $X_1$ and $X_2$ denote independent copies of $X$.
\end{thm}

\vspace{0.1in}

Simply rearranging the conclusion of Theorem~\ref{thm:bn} 
gives the following stability result.

\begin{cor}\label{cor:bn}
If $X$ is a log-concave random vector in $\RL^n$, 
with Poincar\'e constant $c$, then:
\ben
\frac{D(X)}{n} \leq \frac{2(1+c)}{c} \log \sigma_{+}(X).
\een 
\end{cor}


\vspace{0.1in}

\begin{rmk}\label{rmk:lc-gaus}
It was proved in \cite[Proposition V.6]{BM11:it}, by relying on an important result of Klartag \cite{Kla06}, that log-concave distributions are not too far
from Gaussianity, in the sense that,
\ben
\frac{D(X)}{n} \leq \frac{1}{4} \log n + C ,
\een
for some absolute constant $C$. Therefore, the main
value of the result in Corollary~\ref{cor:bn} 
is in that it explicitly connects ``non-Gaussianity'' 
with the doubling constant $\sigma_+(X)$, and especially
when $\sigma_+(X)$ is small. 
\end{rmk}

Interestingly, it is not hard 
to give a much more elementary converse result 
when we know something about both the
doubling and the difference constants.
Indeed, we show below that any random vector whose doubling 
and difference constants differ significantly, must also 
be significantly
far from Gaussianity.

\begin{thm}
\label{thm:gaussC}
If $X_1$ and $X_2$ are independent copies of any random vector $X$ in $\RL^n$ with
finite differential entropy, then,
\ben
\frac{D(X)}{n} \geq \frac{1}{4} |\log \sigma_+(X)-\log \sigma_-(X)| .
\een
\end{thm}

\begin{IEEEproof}
By the invariance of the entropy under linear 
transformations of determinant 1,
\ben
h(X_1)+h(X_2)&=&h(X_1, X_2) \\
&=& h\bigg(\frac{X_1+X_2}{\sqrt{2}}, \frac{X_1-X_2}{\sqrt{2}}\bigg) \\
&\leq&
 h\bigg(\frac{X_1+X_2}{\sqrt{2}}\bigg) + h\bigg(\frac{X_1-X_2}{\sqrt{2}}\bigg) .
\een
Let $a$ be the greater of the quantities $h\big(\frac{X_1+X_2}{\sqrt{2}}\big)$ and $h\big(\frac{X_1-X_2}{\sqrt{2}}\big)$,
and $b$ be the lesser of them. The above display implies that,
\be\label{eq:symm-epi}
\frac{a+b}{2} \geq h(X) .
\ee
Now, by the scaling property of differential entropy, we have,
\ben
\half |h(X_1-X_2)-h(X_1+X_2)|
&=& \half \bigg| h\bigg(\frac{X_1+X_2}{\sqrt{2}}\bigg) - h\bigg(\frac{X_1-X_2}{\sqrt{2}}\bigg) \bigg| \\
&=& \frac{a-b}{2}
= a- \frac{a+b}{2} \\
&\leq& a-h(X) ,
\een
using \eqref{eq:symm-epi} to obtain the inequality. Since both $\frac{X_1+X_2}{\sqrt{2}}$ and $\frac{X_1-X_2}{\sqrt{2}}$
have the same covariance matrix as $X$, the maximum entropy property of the Gaussian implies that
$h(Z)\geq a$, where $Z$ is a Gaussian random vector with the same covariance matrix as $X$.
Thus we have,
\ben
\half |h(X_1-X_2)-h(X_1+X_2)| \leq h(Z)- h(X) = D(X) ,
\een
which is equivalent to the desired statement by using the relations \eqref{eq:ruz-diff} and \eqref{eq:ruz-doub}.
\end{IEEEproof}
\vspace{.1cm}

%



\newpage

\section{An explicit reverse entropy power inequality}
\label{sec:repi}

In recent work \cite{BM11:cras}, a reverse entropy power 
inequality was developed for
the class of log-concave distributions. Recall that the 
entropy power inequality due to Shannon and Stam 
\cite{Sha48,Sta59} asserts that
$\calN(X+Y) \geq \calN(X) + \calN(Y)$,
for any two independent random vectors $X$ and $Y$ in $\RL^n$ for which
the entropy is defined. 
The entropy power inequality may be formally strengthened by using the
invariance of entropy under affine transformations of determinant $\pm 1$, 
i.e., $\calN(u(X)) = \calN(X)$ whenever $|{\rm det}(u)|=1$. Specifically,
\be\label{epi-aff}
\inf_{u_1, u_2} \calN(u_1(X)+u_2(Y)) \geq \calN(X) + \calN(Y),
\ee
where the maps $u_i:\RL^n \rightarrow \RL^n$ range over all 
affine entropy-preserving transformations. 
What \cite{BM11:cras} showed was that the inequality \eqref{epi-aff} 
can be reversed with a constant independent of dimension 
if we restrict to log-concave distributions.

\begin{thm}[{\sc Reverse EPI, \cite{BM11:cras}}]
\label{thm:repi}
If $X$ and $Y$ are independent random vectors 
in $\RL^n$ with log-concave densities, 
there exist linear entropy-preserving 
maps $u_i:\RL^n \rightarrow \RL^n$ such that 
\be\label{eq:repi}
\calN\big(\widetilde X + \widetilde Y\big)\, \leq\, C\, (\calN(X) + \calN(Y)),
\ee
where $\widetilde X = u_1(X)$, $\widetilde Y = u_2(Y)$, and where
$C$ is a universal constant.
\end{thm}
\vspace{.1cm}

This reverse entropy power inequality is analogous
to Milman's \cite{Mil86} 
reverse Brunn-Minkowski inequality 
(see also \cite{Mil88:2,Mil88:1,Pis89:book}), 
which is a celebrated result in convex geometry.
In this light, Theorem~\ref{thm:repi} can be seen
as an extension of the analogies between geometry and information theory
that were previously observed by Dembo, Cover and Thomas 
\cite{DCT91}, among others. Also, Theorem~\ref{thm:repi}
can be extended to the larger subclass of
so-called ``convex measures'' \cite{BM12:jfa}.


Observe that the universal constant provided by the proof 
of Theorem~\ref{thm:repi} is not explicit, and it is not easy 
to even get bounds on it. But in the special case when $X$
and $Y$ have the same distribution, 
we show below 
that an explicit constant
can be obtained rather simply. To do this, we first note that
the following result of Cover and Zhang \cite{CZ94} 
easily generalizes to higher dimensions:
If $X$ and $X'$ are (possibly dependent) random variables
with the same log-concave marginal distribution on $\RL$,
then,
$h(X+X') \leq h(2X)$.

\begin{thm}\label{thm:iid-repi}
If $X$ and $Y$ are (possibly dependent) random vectors in
$\RL^n$, with the same
log-concave marginal density, then, 
\ben
h(X+Y)\leq h(2X) .
\een
\end{thm}

\begin{IEEEproof}
Suppose the common marginal density of $X$ and $Y$ is $f$,
and let $g$ be the density  of $Z=X+Y$.
Since $f$ is log-concave, Jensen's inequality implies that,
\begin{align*}
E\log f\bigg(\frac{X+Y}{2}\bigg) &\geq  E \half [\log f(X) + \log f(Y)] \\
&= \half [ E\log f(X) + E\log f(Y)] \\
&= -h(f) .
\end{align*}
Observe that independence is not required here, and all
expectations are taken with respect to the joint distribution of $(X,Y)$.
In particular, we have that,
\begin{align*}
\int g(z) \log \tilde{f}(z) dz
&= \int g(z)  \log f(\frac{z}{2}) - 1 \\
&\geq -h(f) -1 \\
&= -h(\tilde{f}) ,
\end{align*}
where $\tilde{f}(z)=\half f(z/2)$ is the density of $Z^*=2X$.
In other words,
\ben
D(g\|\tilde{f}) + h(g) \leq h(\tilde{f}) .
\een
Thus $h(g)=h(X+Y)$ is maximized if and only if $g=\tilde{f}$, i.e., 
when $X$ and $Y$ are identical. 
\end{IEEEproof}
\vspace{.1in}

Theorem~\ref{thm:iid-repi} immediately implies that 
for i.i.d. random vectors with log-concave distribution, the 
reverse entropy power inequality (Theorem~\ref{thm:repi}) holds with 
both linear transformations being the identity, and with a universal constant of 2.

\begin{cor}\label{cor:repi-iid}
If $X, X'$ are independent random vectors with
the same
log-concave distribution, then,
\ben
\calN(X+X')\leq 2[\calN(X)+\calN(X')].
\een
In other words, for any  log-concave random vector $X$,
$\sigma_{+}(X)\leq 2$.
\end{cor}

\begin{IEEEproof}
From Theorem~\ref{thm:iid-repi},
\ben
\calN(X+X')\leq \calN(2X)= 4\calN(X) = 2[\calN(X)+\calN(X)].
\een
\end{IEEEproof}
\vspace{.1in}

A version of  Corollary~\ref{cor:repi-iid} was obtained 
(contemporaneously with this work) by a different method in \cite{BM13:goetze};
however the bound on the doubling constant in that work is $e^4/2\approx 27.3$,
which is significantly worse than the bound of 2 we obtain.
Soon after the first version of this paper was released, some related results and
a nice conjecture about reverse forms of the entropy power inequality were
released by Ball, Nayar and Tkocz \cite{BNT15}.

Although it already seems rather restrictive that the doubling constant of
any log-concave random vector lies between 1 and 2, we do not believe the upper 
bound is optimal. However, Corollary~\ref{cor:repi-iid} represents
yet another way in which general log-concave random vectors resemble Gaussian
ones; as mentioned in Remark~\ref{rmk:lc-gaus}, \cite{BM11:it} gives a different
formulation of this intuition.

Another way to view Corollary~\ref{cor:repi-iid} is in the context of the central limit theorem.
Recall that the central limit theorem in terms of
relative entropy (\cite{Bar86, ABBN04:1}, see also \cite{MB07}) 
asserts that if $X, X_1, X_2, \ldots$ are i.i.d.
random vectors with $h(X)>-\infty$, then, as $n\to\infty$,
\ben
h\bigg(\frac{X_1+\ldots+X_n}{\sqrt{n}}\bigg) \uparrow h(N(0,I_n) .
\een
Corollary~\ref{cor:repi-iid}  implies that,
\ben
\calN(X) \leq \calN\bigg(\frac{X_1+X_2}{\sqrt{2}}\bigg)\leq 2\calN(X) ,
\een
and hence constrains the rate at which entropy can increase when doubling sample size 
in the central limit theorem for i.i.d. log-concave summands. 

The above development is also closely related to a very nice observation of K.~Ball, dating back
to around 2003 but with details only being published much later in \cite{BN12}, 
relating two important conjectures in convex geometry, namely the Kannan-Lov\'asz-Simonovits
conjecture \cite{KLS95} and the hyperplane conjecture or slicing problem of Bourgain \cite{Bou86}.
We explain this connection in our language; the reasoning is related to that of
K.~Ball even if it differs in details. The Kannan-Lov\'asz-Simonovits (KLS) conjecture  asserts that 
the Poincar\'e constant $c$ is bounded from below 
for all log-concave densities by a universal
constant $C$ independent of dimension.
If this is true, then Corollary~\ref{cor:bn} 
implies that 
\ben
\frac{D(X)}{n} \leq 2 \bigg(1+\frac{1}{C}\bigg) \sigma_{+}(X) \leq 2 \bigg(1+\frac{1}{C}\bigg) \log 2 ,
\een
using Corollary~\ref{cor:repi-iid} for the second inequality. In other words,
$D(X)/n$ is bounded by a universal constant for any log-concave random vector $X$ in $\RL^n$,
which by \cite[Corollary 5.3]{BM11:it}, is equivalent to the hyperplane conjecture
(whose original formulation in \cite{Bou86} we do not bother to state here). 
Hence the KLS conjecture implies the hyperplane conjecture.



\newpage

\section{Towards a Rogers-Shephard inequality for entropy}
\label{sec:rs}

The Rogers-Shephard inequality \cite{RS57} asserts that,
if $K\subset \RL^n$ is a convex body, then
\be\label{rs}
\vol(K-K) \leq \binom{2n}{n} \vol(K) ,
\ee
with equality if and only if $K$ is the $n$-dimensional simplex.
It complements the fact, implied by the Brunn-Minkowski
inequality, that,
\be\label{bm-spl}
\vol(K-K) \geq 2^n \vol(K) .
\ee
Indeed, since by Stirling's formula and some algebraic manipulation,
\ben
\binom{2n}{n} < 4^n,
\een
the inequalities \eqref{rs} and \eqref{bm-spl} together imply,
\ben
2 \vol(K)^{1/n} \leq \vol(K-K)^{1/n} < 4 \vol(K)^{1/n} . 
\een

As suggested by the analogy between the reverse entropy power inequality
and the reverse Brunn-Minkowski inequality discussed in the preceding section,
the natural probabilistic analogue of a convex set is a log-concave distribution,
and a natural probabilistic analogue of volume is entropy. Therefore, it is 
natural to ask whether there is a probabilistic analogue of the Rogers-Shephard inequality. Indeed, we show that for 
$X, X'$ i.i.d. log-concave random vectors, $\calN(X-X')$ is bounded
by a multiple of $\calN(X)$.


\begin{cor}\label{cor:rs-ent}
If $X, X'$ are independent random vectors drawn from the same
log-concave distribution, then
\ben
\calN(X-X')\leq 16 \calN(X) .
\een
In other words, for any  log-concave random vector $X$,
$\sigma_{-}(X)\leq 8$.
\end{cor}

\begin{IEEEproof}
By Corollary~\ref{cor:repi-iid}, 
\ben
\calN(X+X')\leq 4\calN(X),
\een
and by Corollary~\ref{cor:sd-iid},
\ben
\calN(X-X')\leq \frac{\calN^2(X+X')}{\calN(X)} \leq 16 \calN(X).
\een
\end{IEEEproof}
\vspace{.1in}

Corollary~\ref{cor:rs-ent} does not provide a tight bound. Indeed, in the contemporaneous work \cite{BM13:goetze},
a different approach is used to obtain a bound on the difference constant of $e^2/2\approx 3.7$, which is better than
our bound of 8. We state below a conjecture for the sharp constant in the one-dimensional case.

\begin{conj}
If $X, X'$ are independent $\RL$-valued random variables drawn from the same
log-concave distribution, then,
\ben
\calN(X-X')\leq 4 \calN(X) ,
\een
with equality if and only if $X$ is a translated and scaled version of the (one-sided) exponential distribution. 
In other words, for any  log-concave random variable $X$,
$\sigma_{-}(X)\leq 2$.
\end{conj}



Of course, we may also write Corollary~\ref{cor:rs-ent} and
Corollary~\ref{cor:repi-iid} in terms of the Ruzsa divergence
using the identities \eqref{eq:ruz-diff} and \eqref{eq:ruz-doub}.

\begin{cor}\label{cor:ruz-div-ub}
If $X$ is a log-concave random vector taking values in $\RL^n$,
then,
\ben
d_R(X\|X)\leq 2n \log 2
\quad\text{and}\quad
d_R(X\|-X)\leq n \log 2 .
\een
\end{cor}

Let us note that a sharp functional analogue of the Rogers-Shephard inequality has been proved
by Colesanti \cite{Col06} for log-concave functions as opposed to densities (see also \cite{AEFO14, AGJV14}).


\newpage

\section{Determinant inequalities}
\label{sec:det}


Differential entropy inequalities have been used to 
to deduce inequalities for positive-definite matrices since
Cover and El Gamal's work in \cite{CG83}; see also \cite{DCT91} and 
\cite{MT10}.
However, in most of these cases, the inequalities deduced relate
determinants of a positive-definite matrix to those of its square submatrices.
We discuss below the use of differential entropy inequalities to prove
determinantal inequalities for sums of positive-definite matrices.
As in the above papers, the main idea is to use the fact that,
for the Gaussian distribution on $\RL^n$ with covariance matrix $K$, 
written $\gamma_K=N(0,K)$, the differential entropy is given by,
\ben
h(\gamma_K)= \half\log \bigg[(2\pi e)^{n} \det(K)\bigg].
\een

A classical inequality for the determinant of sums is Minkowski's
inequality, which asserts that,
for $n\times n$ positive-definite matrices,
\ben
\det(A+B)^{\nth} \geq \det(A)^\nth + \det(B)^\nth .
\een
This may be seen as a consequence of the entropy power inequality
(by specializing to Gaussians), but there are also elementary means
of deriving it.

On the other hand, upper bounds for the determinant of a sum of 
positive-definite matrices are not as well known. This is 
partly due to the fact that
the most straightforward inequalities that one might try to check, 
like subadditivity, are actually false. However, Rotfel'd \cite{Rot67} 
did obtain
such a bound when one of the matrices involved is the identity matrix:
\be\label{eq:rot}
\det(I+A+B) \leq \det(I+A) \cdot \det(I+B) .
\ee
Indeed, he obtained this as a special case of a more general inequality
for arbitrary square matrices,
\ben
\det(I+|A+B|) |\leq \det(I+|A| ) \cdot \det(I+|B| ) ,
\een
where $|A|=\sqrt{A^* A}$ and $A^*$ is the adjoint of $A$.


Our final observation is that, 
substituting normals in Theorem~\ref{thm:p-r},
provides an extremely simple alternative proof 
of a generalization of inequality \eqref{eq:rot},
not requiring any of the matrices to be the identity:

\begin{cor}
\label{cor:det}
Let $K$ and $K_{i}$ be positive-definite matrices of the same dimension.
Then:
\ben
\det(K +K_{1}+\ldots+K_{n}) \leq [\det(K)]^{-(n-1)} 
\prod_{j=1}^n \det(K+ K_{j}) .
\een
\end{cor}





\newpage



\end{document}